\pdfoutput=1

\documentclass[sigconf]{acmart}
\acmConference[ASE 2021]{The 36th IEEE/ACM International Conference on Automated Software Engineering}{15 - 19 November, 2021}{Melbourne, Australia}
\usepackage[labelfont=bf,textfont={bf}]{caption}

\usepackage{dblfloatfix}
\usepackage{arydshln}

\usepackage{balance}
\usepackage{alltt}
\usepackage{subcaption}
\usepackage{graphicx}
\usepackage{colortbl}
\usepackage{tabularx}
\usepackage{tabu}
\usepackage{array}

\usepackage{booktabs}
\usepackage{wrapfig}
\usepackage{tikz}
\usetikzlibrary{angles}
\usepackage{multirow}
\usepackage{hyperref}
\usepackage{framed}
\usepackage[framemethod=tikz]{mdframed}
\usetikzlibrary{shadows}
\usepackage{enumitem}
\usepackage{graphics}
\usepackage{cleveref}

\setlist[description]{leftmargin=1cm}

\usepackage{amsmath}

\usepackage{pgfplots}
\usetikzlibrary{patterns}
\usetikzlibrary{arrows}
\usetikzlibrary {positioning}
\usepackage{enumitem}
\setlist[itemize]{leftmargin=*}
\setlist[enumerate]{leftmargin=*}
\usepackage{makecell}
\usepackage{xcolor,pifont}
\usepackage[linesnumbered,ruled,vlined]{algorithm2e}
\usepackage{natbib}
\setlist{nolistsep} 

\SetKwProg{Fn}{Function}{}{}

\setlist[1]{itemsep=0pt}

\newmdenv[
tikzsetting= {fill=gray!10},
linewidth=1pt,
roundcorner=2pt,
shadow=false
]{myshadowbox}

\newcolumntype{P}[1]{>{\centering\arraybackslash}p{#1}}

\SetKwInput{KwInput}{Input}                
\SetKwInput{KwOutput}{Output}              

\SetKwFunction{FMain}{Main}
\SetKwFunction{FSum}{Sum}
\SetKwFunction{FSub}{get\_ngbr}

\hypersetup{
    linkcolor=blue,
    filecolor=magenta,      
    urlcolor=cyan,
}

\newcommand{\bi}{\begin{itemize}[leftmargin=0.4cm]}
\newcommand{\ei}{\end{itemize}}
\newcommand{\be}{\begin{enumerate}[leftmargin=0.4cm]}
\newcommand{\ee}{\end{enumerate}}

\usepackage{url}

\tikzstyle{thmbox} = [rectangle, rounded corners, draw=black, fill=gray!10]

\usepackage{xcolor}
\usepackage[tikz]{bclogo}
\newenvironment{RQ}[1]%
{\noindent\begin{minipage}[c]{\linewidth}%
\begin{bclogo}[couleur=gray!25,%
                arrondi=0.1,%
                logo=\bctrombone,%
                ombre=true]{{\normalsize ~#1}}}%
{\end{bclogo}\end{minipage}\vspace{2mm}}

\newcommand*{\rowstyle}[1]{
  \gdef\@rowstyle{#1}%
  \@rowstyle\ignorespaces%
}

\newcommand*\colourcheck[1]{%
  \expandafter\newcommand\csname #1check\endcsname{\textcolor{#1}{\ding{51}}}%
}

\colourcheck{green}

\newcommand*\colourx[1]{%
  \expandafter\newcommand\csname #1x\endcsname{\textcolor{#1}{\ding{53}}}%
}
\colourx{red}

\usepackage{listings}
\usepackage{cellspace}
\newcommand{\revised}{\textcolor{black}}

\title{  FRUGAL:
 Unlocking Semi-Supervised Learning \\for Software Analytics }

\renewcommand{\shorttitle}{FRUGAL:
Unlocking SSL for Software Analytics}

\begin{document}




\author{Huy Tu, 
          Tim Menzies\\
          Com Sci, NCState, USA\\
         hqtu@ncsu.edu, timm@ieee.org
 }


\begin{abstract}
Standard software analytics often involves having a large amount of data with labels in order to commission models with acceptable performance. However, prior work has shown that such requirements can be expensive, taking several weeks to label thousands of commits, and not always available when traversing new research problems and domains. Unsupervised Learning is a promising direction to learn hidden patterns within unlabelled data, which has only been extensively studied in defect prediction. Nevertheless, unsupervised learning can be ineffective by itself and has not been explored in other domains (e.g.,  static analysis and issue close time).

Motivated by this literature gap and technical limitations, we present FRUGAL, a tuned semi-supervised method that builds on a simple optimization scheme that does not require sophisticated (e.g., deep learners) and expensive (e.g., 100\% manually labelled data) methods. FRUGAL  optimizes the unsupervised learner's configurations (via a simple  grid search) while validating our design decision
of labelling just \revised{2.5\%} of the   data before prediction. 

As shown by the experiments of this paper FRUGAL outperforms the state-of-the-art adoptable
static code warning recognizer and issue closed time predictor, while reducing the cost of labelling by \revised{a factor of 40 (from 100\% to 2.5\%)}.
Hence we assert that FRUGAL can save considerable effort
in data labelling especially in validating prior work or researching new problems. 

Based on this work, we suggest
 that proponents of complex and expensive methods should always baseline such methods against simpler and cheaper alternatives. For instance, a semi-supervised learner like FRUGAL can serve  as a baseline to the state-of-the-art software analytics.



\end{abstract}

\maketitle

\section{Introduction}\label{sec:introduction}

Software analytics can guide improvements to software quality, maintenance and security. For example, analytics can discover which static code warnings are adoptable~\cite{intrinsic_static, wang2018there}; whether the new issues can be easily fixed~\cite{simple_ict, mani2019deeptriage}; where software defects are likely to occur~\cite{agrawal2019dodge, Ostrand:2004};  which comments likely to contain technical debts~\cite{maldonado2015detecting, jitterbug}; what the current health conditions of these open-source projects~\cite{projecthealth_xia}; or how to distinguish security bug reports~\cite{rui_security}.

However, models that perform these software analytics tasks typically learn from  {\em
labelled data}. Generating such labels can be extremely slow and expensive.
For instance,~\citet{tu2020better} reported that manually reading
and labelling $22,500+$ commits 
required 175 person-hours (approximately nine weeks), including cross-checking among labellers.
Due to the labor-intensive nature of the process, researchers often reuse
datasets labelled from previous studies. For instance, Lo et al.~\cite{yang2015deep}, Yang et al.~\cite{yang16unsupervised}, and Xia et al.~\cite{xia16ist17} certified their methods using data generated by Kamei et al.~\cite{kamei12_jit}. While this practice allows researchers to rapidly
test new methods, it leaves the possibility for  
any labelling mistake
to propagate to other related works. In fact, in technical debts identification, before reusing prior work's data~\cite{maldonado2015detecting},  Yu et al.~\cite{jitterbug} discovered that more than 98\% of the false positives were actually true positives, casting doubt on work that used the original dataset.
Hence, it is timely to ask:
\begin{quote}
\begin{center}

{\em
Can we reduce the labelling effort associated
with building models for  software analytics?
}
 
\end{center}
\end{quote}
Unsupervised learning techniques that learns patterns from unlabelled data is a promising direction for software analytics. 
Such learning has been used for buggy/non-buggy classification~\cite{yang16unsupervised,unsup_review,yang2016effort,yan2017file,zhang2016cross,yang2016defect}. The state-of-the-art (SOTA) unsupervised learner is~\citet{nam2015clami}'s CLA($C$) method. CLA is based on the binary split of the output space at the aggregated median ($C=50\%$) of all features' median in the data. However, other areas and different datasets may not share the same data characteristics for the default CLA ($C=50\%$) to perform well. To address this gap, our study adopts and extends CLA from defect prediction to other software analytics like static code warnings and issues close time. 

\begin{figure*}[!t]
 
\centering \includegraphics[width=0.95\linewidth]{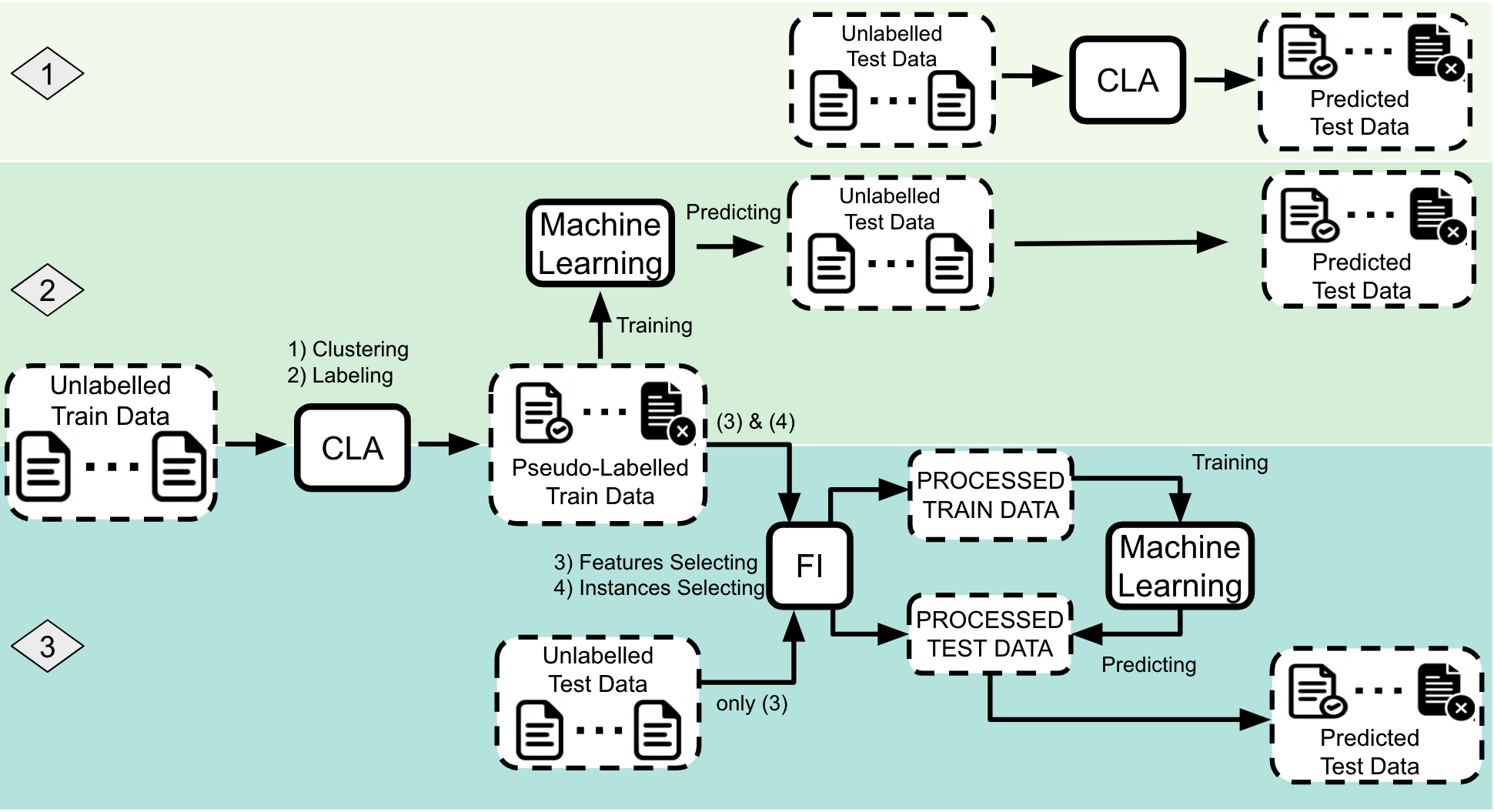}
\vspace{-10pt}
\caption{Three different modes of CLA devised from \citet{nam2015clami} for defect prediction.}
 
\label{fig:cla}
\end{figure*}

Promising extensions for unsupervised learning involves finding different control settings to configure the system (hyperparameter tuning) and validating on small labelled data regions (e.g., \revised{2.5\%}) before applying the best setting to the test data. Recent software engineering (SE) research shows many domains' SOTA can be improved with  hyperparameter tuning~\cite{agrawal2019dodge, di18_fft, Fu17easy, fu16, agrawal2018better, nair2017flash}. DODGE is one prominent optimizer which shows the output space of the models on low-dimensional data can be easily surveyed through dodging away from (1) prior options or options that resulted (2) in statistically similar performance. Simply, the central function of CLA (binary split of the output space via aggregated $C$) is synonymous to SE's SOTA optimizer DODGE with less information required, a reduction of \revised{97.5\%} train data's labels. Specifically, our work proposed FRUGAL($C$) = three different modes of CLA (as shown in Figure \ref{fig:cla}) with $C=\{5\%$ to $95\%$, increments by $5\%\}$ where all the combinations can be easily executed in a grid search manner.

To understand and validate the FRUGAL system, we investigate the following research questions: \\


\noindent \textbf{RQ1: How much labelled data ($L$\%) that FRUGAL requires?}

\begin{RQ}{\normalsize{Result:}} 
From our investigation of various $L$ values, FRUGAL's performance plateaus beyond \revised{$L\geq2.5\%$} and FRUGAL's success is not altered by large changes to $L$.  
\end{RQ}

\noindent \textbf{RQ2: How does FRUGAL perform in adoptable static code warnings identification?} 

\begin{RQ}{\normalsize{Result:}} 
When comparing to the SOTA solution in EMSE'20 \cite{intrinsic_static}, FRUGAL wins on recall, loses in AUC, and draws in FAR with only \revised{2.5\%} of the labelled train data. 
\end{RQ}

\noindent \textbf{RQ3: How well
 does FRUGAL predict    issue close time?}  

\begin{RQ}{\normalsize{Result:}} 
When comparing to the SOTA solution in EMSE'20 \cite{simple_ict} (which was compared to  ICSE'10~\cite{giger2010predicting}, PROMISE'11~\cite{marks2011studying}, MSR'16~\cite{Kikas16},  COMAD'19~\cite{mani2019deeptriage}), FRUGAL outperforms in FAR, recall, and AUC while performing similarly in accuracy with only \revised{2.5\%} of the labelled train data.
\end{RQ}

In summary, our work's contributions to the field of software analytics are as follows:

\be
\item This work is the first to assess the usage of unsupervised learning to reduce the labelling efforts to commission models building in adoptable static code warnings identification and issues close time prediction. 

\item FRUGAL surpasses the SOTA issues close time predictor and performs similarly to the SOTA adoptable static code warning identifier with \revised{97.5\%} less information. 

\item FRUGAL reduces the labelling efforts to commission new models building by \revised{97.5\%}. In another word, FRUGAL is \revised{40} times cheaper than SOTA methods in issue close time and static code warning analysis areas. 

\item The performance of our framework suggests that many more domains in SE could benefit from unsupervised learning \revised{solutions in the semi-supervised learning manner} beyond defect prediction~\cite{nam2015clami,yang16unsupervised,fu2017unsupervised,unsup_review,yang2016effort,yan2017file,9115238,zhang2016cross,yang2016defect,Zhou2018HowFW}. 

\item To better support other researchers our scripts and data are on-line at https://github.com/SE-Efforts/SE\_SSL.

\ee

The rest of this paper is structured as follows. \textit{Section 2} and \textit{3} discusses the background, related works, and motivation of this work. \textit{Section 4} describes our methodologies.  \textit{Section 5} analyzes the results while  \textit{Section 6} discusses our short-comings. Finally, \textit{7} concludes the work  and  states the venues for future work.

\section{Background and Related Work} \label{background}

\subsection{Studying Static Code Warnings}

\subsubsection{Background}

Static code warning tools detect potential static code defects in source code or executable files at the stage of software product development. This covers a range of potential defects such as common
programming errors, code styling, in-line comments common programming
anti-patterns, style violations, and questionable coding decisions.  The distinguishing feature of these tools is that they make their comments without reference to a particular input. Nor do they use feedback from any execution of the code being studied.
Examples of these tools include PMD\footnote{\url{https://pmd.github.io/latest/index.html}}, Checkstyle\footnote{\url{https://checkstyle.sourceforge.io/}} and the  FindBugs\footnote{\url{http://findbugs.sourceforge.net}} tool.
\begin{table}[!htb]
\footnotesize
\vspace{-5pt}
\caption{Categories of Wang et al.~\cite{wang2018there}'s selected features.~(8 categories are shown in the left column, and 95 features explored in Wang et al. are shown in the right column with 23 golden features in bold.)}
\vspace{-5pt}
\tabcolsep=0.2cm
\begin{center}
\begin{tabular}{ll}
\hline
\textbf{Category}  &   \textbf{Features} \\ \hline
\multirow{1}{*}{\rotatebox[origin=c]{0}{\parbox[c]{1cm}{ Warning  \\ Combination}}}  & \begin{tabular}[c]{@{}l@{}}size content for warning type;\\ size context in method, file, package;\\ \textbf{warning context in method, file,} package;\\ \textbf{warning context for warning type};\\ fix, non-fix change removal rate;\\ \textbf{defect likelihood for warning pattern};\\ variance of likelihood;\\ defect likelihood for warning type;\\ \textbf{discretization of defect likelihood}; \\ \textbf{average lifetime for warning type};\end{tabular} \\ \hline
\multirow{1}{*}{\rotatebox[origin=c]{0}{\parbox[c]{1cm}{ Code  \\ characteristics}}}   & \begin{tabular}[c]{@{}l@{}}method, file, package size;\\ comment length;\\ \textbf{comment-code ratio};\\ \textbf{method, file depth};\\ method callers, callees;\\ \textbf{methods in file}, package;\\ classes in file, \textbf{package};\\ indentation;\\ complexity;\end{tabular} \\ \hline
\multirow{1}{*}{\rotatebox[origin=c]{0}{\parbox[c]{0cm}{ Warning\\ characteristics}}}  & \begin{tabular}[c]{@{}l@{}}\textbf{warning pattern;} \\ 
\textbf{type, priority};  \\ rank, warnings in method, file, \\  \textbf{package};\end{tabular} \\ \hline
\multirow{1}{*}{\rotatebox[origin=c]{0}{\parbox[c]{1cm}{ File  \\ history}}}    & \begin{tabular}[c]{@{}l@{}}latest file, package modification;\\ file, package staleness;\\ \textbf{file age}; \textbf{file creation};\\ deletion revision; \textbf{developers};\end{tabular} \\ \hline
\multirow{1}{*}{\rotatebox[origin=c]{0}{\parbox[c]{1cm}{ Code  \\ analysis}}}   & \begin{tabular}[c]{@{}l@{}}call name, class, \textbf{parameter signature},\\ return type; \\ new type, new concrete type;\\operator;\\ field access class, field; \\catch;\\ field name, type, visibility, is static/final;\\ \textbf{method visibility}, return type,\\ is static/ final/ abstract/ protected;\\ class visibility, \\ is abstract / interfact / array class;\end{tabular} \\ \hline
\multirow{1}{*}{\rotatebox[origin=c]{0}{\parbox[c]{1cm}{ Code  \\ history}}}   
 & \begin{tabular}[c]{@{}l@{}}added, changed, deleted, growth, total, percentage \\ of LOC in file in the past 3 months;\\ \textbf{added}, changed, deleted, growth, total, percentage \\ of LOC in file in the last 25 revisions;\\ \textbf{added}, changed, deleted, growth, total, percentage \\ of LOC in package in the past 3 months;\\ added, changed, deleted, growth, total, percentage \\ of LOC in package in the last 25 revisions;\end{tabular} \\ \hline
\multirow{1}{*}{\rotatebox[origin=c]{0}{\parbox[c]{1cm}{ Warning  \\ history}}}   & \begin{tabular}[c]{@{}l@{}}warning modifications;\\ warning open revision;\\ \textbf{warning lifetime by revision}, by time;\end{tabular} \\ \hline
\multirow{1}{*}{\rotatebox[origin=c]{0}{\parbox[c]{2cm}{ File  \\ characteristics}}} & \begin{tabular}[c]{@{}l@{}}file type;\\ file name; \\package name;\end{tabular} \\ \hline
\end{tabular}
\end{center}
\label{table:variables} 
\vspace{-10pt}
\end{table}

One issue with static code warnings is that they generate a large number
of false positives. Many programmers routinely ignore most of the static code
warnings, finding them irrelevant or spurious\revised{~\cite{wang2018there}}. Such warnings are considered as ``unadoptable'' since programmers just ignored them.  Between
35\% and 91\% of the warnings generated from static analysis tools are known
to be unadoptable. This high false alarm rate is one of the most significant barriers for developers to use these tools~\cite{thung2015extent,avgustinov2015tracking,johnson2013don}.  Hence it is prudent to learn to recognize what kinds
of warnings programmers usually act upon so the tools can be made more useful by first pruning away the unadoptable warnings. Various approaches have been tried to reduce these false alarms including graph theory~\cite{boogerd2008assessing,bhattacharya2012graph}, statistical models~\cite{chen2005novel}, and ranking schemes~\cite{kremenek2004correlation}. \revised{Previous work~\cite{intrinsic_static} referred to the target warnings found by these approaches as ``actionable'' warnings, but we found that it actually refer to ``adoptable'' warnings that were adopted. That means the warnings that are adopted by developers do not necessarily mean the warnings are actionable (some developers still need to consult external sources to figure out the solutions).}

\subsubsection{Data and Algorithms}

The data for this paper comes from a recent study by Wang et al.~\cite{wang2018there}.
They conducted a systematic literature review to collect all public available static code features generated by widely-used static code warning tools (116 in total):


\bi

\item
All the values of these collected features were extracted from warning reports generated by FindBugs based on 60 revisions of 12 projects.

\item To ensure the difference between prior and later revision intervals of a project is adequate for the solid conclusions to be drawn, Wang et al.~\cite{wang2018there} set revision intervals for different projects, e.g., 3 months for \textit{Lucene} and 6 months for \textit{Mvn}. Each project in this study has at least two-years commit history.
\item To eliminate ineffective features to the results of those learners, a greedy backward selection algorithm is applied. Then they isolated 23 features as the most useful ones for identifying adoptable static code warnings. 
\item
They called these features the ``golden set''; i.e. the features most important for recognizing adoptable static code warnings.
\ei
To the best of our knowledge, this is the most exhaustive research about static warning characteristics yet published.
As shown in Table~\ref{table:variables},
the  ``golden set'' features   fall into eight categories.
These features are the independent variables used in this study. To assign dependent labels, we applied the methods of
Liang et al.~\cite{liang2010automatic}.
They defined a specific warning as adoptable if it is closed after the later revision interval.
\begin{table}[!t]
\vspace{-5pt}
\caption{Summary of \citet{intrinsic_static}'s data distribution. The gray cells are median values for the corresponding columns.}
\vspace{-13pt}
\begin{center}
\resizebox{\linewidth}{!}{\begin{tabular}{@{}rrrrrrr@{}}
\midrule
 &&  & \multicolumn{2}{c}{\textbf{training set}} & \multicolumn{2}{c}{\textbf{test set}} \\ \midrule
& \textbf{Dataset} &  \textbf{Features}& 
 \begin{tabular}[c]{@{}c@{}}\textbf{Instance} \\ \textbf{Counts}\end{tabular} & \begin{tabular}[c]{@{}c@{}}\textbf{Adoptable} \\ \textbf{Ratio(\%)}\end{tabular} & \begin{tabular}[c]{@{}c@{}}\textbf{Instance} \\ \textbf{Counts}\end{tabular} & \begin{tabular}[c]{@{}c@{}}\textbf{Adoptable} \\ \textbf{Ratio(\%)}\end{tabular} \\ \midrule
 
 & commons &39& 725 & 7 & 786 & 5 \\ 
 & phoenix &44&  2235 & 18 & 2389 & 14 \\
 & mvn (maven) &47& 813 & 8 & 818 & 3 \\
 & jmeter &49& 604 & 25 & 613 & 24 \\
 & cass (cassandra) &\cellcolor{gray!20}55& 2584 & \cellcolor{gray!20}15 & 2601 & \cellcolor{gray!20}14 \\
 & ant &56& 1229 & 19 & 1115 & 5 \\
 & lucence &57& 3259 & 37 & 3425 & 34 \\
 
 & derby &58& 2479 & 9 & 2507 & 5 \\
 
 & tomcat &60& \cellcolor{gray!20}1435 & 28 & \cellcolor{gray!20}1441 & 23 \\
 \bottomrule
\end{tabular}}
\end{center}
\vspace{-17pt}

\label{table:distribution}
\end{table}

By analyzing FindBugs output from two consecutive releases of nine software projects, collecting the features of Table~\ref{table:variables}, and then applying the Liang et al.'s definitions, we created the data of
Table \ref{table:distribution}. 
In this table, the ``training set'' refers to release  $i-1$ and the ``test set'' is release $i$. In this study, we only employ two latest releases. One of many extensive studies exploring the usage of Machine Learning (ML) in this area is Heckaman et al.~\cite{heckman2009model}. They applied 15 ML algorithms to recognize the adoptable warnings (programmers can act upon) based on 51 features derived from static analysis tool,
they achieved  recalls of  83-99 \% (average across 15 data sets). The SOTA system that we will compare against is from \citet{intrinsic_static} where they took advice from Ghotra et al.~\cite{ghotra15} to compare several representative non-neural learners (Table 9 of \cite{ghotra15}) in software analytics with various popular neural-network models. They found that all treatments performed similarly to each other but non-neural learners did that with less time than deep learners.

Note that, for any particular data set,
the 23 \revised{features} of  Table~\ref{table:variables},
can grow to more than 23 features. For example, consider the ``return type'' feature in the ``code analysis'' category. This can include numerous return types extracted from a given project, which could be void, int, URL, boolean, string, printStream, file, and date~(or a list of any of these periods).
Hence, as shown in Table \ref{table:distribution},
the number of features in our data varied from 39 to 60.



\subsection{Predicting Bugzilla Issue Close Time}\label{sec:tasks}

\subsubsection{Background}

When programmers work on repositories, predicting  issue  close  time  has  multiple  benefits  for  the  developers,  managers, and stakeholders since it is helpful for (1) end-users who are directly affected by the product; (2) developers prioritize work; (3) managers allocate resources and improve consistency of release cycles;  and (4) stakeholders understand changes in project timelines and budgets:
\begin{itemize}
\item
Although bugs have an assigned severity, this is not a sufficient predictor for the lifetime of the issue. For example, the author who issued the bug may be significant contributors to the project.
\item
Alternatively, an issue deemed more \textit{visible} to end-users may be given higher priorities. It is therefore insufficient simply to consider the properties of the issue itself (the \textit{issue metrics}), but also of its environment ( \textit{context metrics}). This is similar to the recent work on how \textit{process metrics} are better defect predicting measurements than \textit{product metrics} \cite{majumder2020revisiting}.
\end{itemize}

An example of such issue close times estimator can notify involved parties if the recently created issue is an easy fix.


%
\subsubsection{Data and Algorithms}

The state-of-the-art system for predicting issue close time comes from a recent study by \citet{simple_ict}. They conducted a literature review of 99 research papers that are comprised of (1) from Watson's literature reviews; and (2) top venues listed in Google Scholar metrics for Software Systems, Artificial Intelligence, and Computational Linguistics in the last three years with at least 10 citations per years:
\bi
\item Traditional or non-neural approaches include (1) \citet{guo2010characterizing}'s study on a large closed-source project (Microsoft Windows) to predict whether or not a bug will be fixed; and (2) \citet{marks2011studying} used ensemble method of decision trees, i.e., random forests, on Eclipse and Mozilla data.
\item As to deep learning or neural network approach are DASENet~\cite{lee2020continual} and DeepTriage\cite{mani2019deeptriage}.

\item Only a minority of deep learning papers (39.4\%)  performed any sort of hyper-parameter optimization, i.e., varied few numbers of parameters, such as the number of layers of the deep learner, to edge out the best performance of deep learning. Even fewer papers (18.2\%) applied hyper-parameter optimization in a non-trivial manner;  i.e., not  using a hold-out set to assess the tuning before assessing the separate test set). 

\ei


To obtain a fair comparison with the prior state-of-the-art, we use the same data as used
in the  \citet{lee2020continual, mani2019deeptriage, simple_ict}'s studies. The data was collected from  the three projects of Firefox, Chromium, and Eclipse: 

\bi
\item Preprocessing involves standard text mining to remove special characters or stack traces, tokenization, and pruning the corpus to a fixed length. 
\item The activities per day were collected into two bins including user activity (e.g., comments), system records (e.g., added/removed labels), and metadata (e.g., the user was the reporter, days from opening, etc). \revised{Given issue close times (1, 2, ... k days), they are grouped into S1 and S2 such that $|S1| \approx |S2|$ and $S1 = \{1, ... i\}, S2 = \{i+1,...,k\}$. For instance, S1 includes 1-43 days and S2 includes 44-365 days, if the number of issues closed in 1 to 43 days $\approx$ number of issues closed in 43 to 365 days.}

\item Along with the numerical metadata, user and system records are transformed to machine-readable data for the models to execute through word2vec \cite{mikolov2013efficient, mikolov2013distributed}. 
\ei

\begin{table}[!t]
\vspace{-5pt}
    \centering
    \caption{An overview of the data used in the \citet{lee2020continual}, \citet{mani2019deeptriage}, and \citet{simple_ict} studies. Note that because of the manner of data collection, i.e., using bin-sequences for each day for each report, there are many more data samples generated from the number of reports mined.}
    \vspace{-5pt}
    \resizebox{\linewidth}{!}{\begin{tabular}{llccc}
    \toprule 
    \textbf{Project} & \textbf{Observation Period} & \textbf{\# Reports} & \textbf{\# Train} & \textbf{\# Test} \\ \midrule
        Eclipse & Jan 2010--Mar 2016 & 16,575 & 44,545 & 25,459 \\
        Chromium & Mar 2014--Aug 2015 & 15,170 & 44,801 & 25,200 \\
        Firefox & Apr 2014--May 2016 & 13,619 & 44,800 & 25,201 \\
        \bottomrule
    \end{tabular}}
    
    \label{tab:bugzilla-data}
    \vspace{-15pt}
\end{table}

In the same manner as prior work, the target class is discretized into two bins (so that each bin has roughly the same number of samples). This yields datasets that are near-perfectly balanced (e.g., in the Chromium dataset, we observed a 49\%-51\% class ratio).

\subsection{Evaluation}

\begin{figure*}[!t]
\vspace{-10pt}
\centering \includegraphics[width=\linewidth]{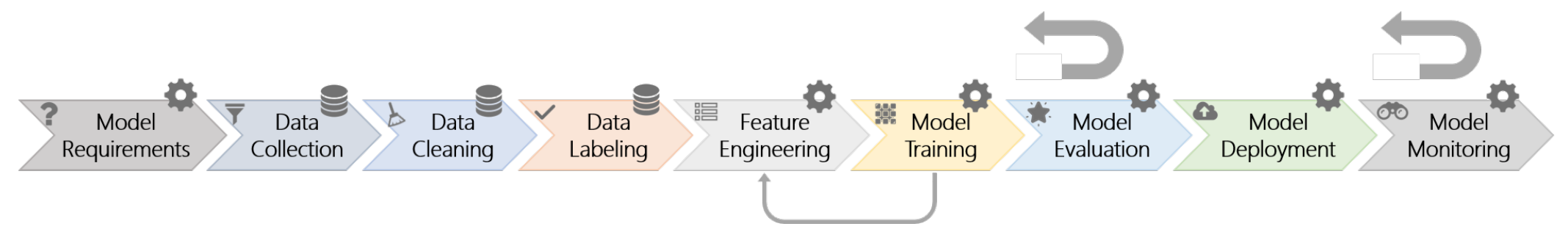}
\vspace{-20pt}
\caption{Nine stages of the machine learning workflow from a case study at Microsoft by \revised{Amershi} et al. \cite{amershi_ICSE19AI4SE}. Some stages are data-oriented (e.g., data collection, cleaning, and labelling) and others are model-oriented
(e.g., model requirements, features engineering, model training, evaluation, evaluation, deployment and monitoring).}\label{fig:pipeline}
\vspace{-5pt}
\end{figure*}   

\subsubsection{Measures of Performance}\label{measures}

Since we wish to compare our approach to prior work, we take the methodological step of adopting the same performance scores as that seen in prior work. Let TP, TN, FP, FN are the true positives, true negatives, false positives, and false negatives (respectively), then \citet{intrinsic_static} used AUC, recall, and false-alarm while \citet{simple_ict} using only accuracy for their studies. \revised{We also add precision and f1 for additional validation (but see our cautionary note
at the end of this list):
 }



\begin{itemize}
    \item \textbf{AUC} (Area Under the ROC Curve) measures the two-dimensional area under the Receiver Operator Characteristic (ROC) curve~\cite{witten2016data,heckman2011systematic}. It provides an aggregate and overall evaluation of performance across all possible classification thresholds to overall report the discrimination of a classifier~\cite{wang2018there}. 
    \item \revised{\textbf{Precision} = $TP/(TP+FP)$ represents the ability of one algorithm to identify instances of positive class among the retrieved positive instances. 
}
    \item \textbf{Recall} = $TP/(TP+FN)$ represents the ability of one algorithm to identify instances of positive class from the given dataset.
    \item \revised{\textbf{F1} = $(2 * Precision * Recall)/(Precision + Recall)$ is the harmonic mean of both precision and recall metrics.}
    \item \textbf{False Alarms (FAR)} = $TN/(TN+FP)$ measures the instances that are falsely classified by an algorithm as positive which are actually negative. This is an important index used to measure the efficiency of a model.
    \item \textbf{Accuracy} = $\mathit{(TP+TN)/(TP+TN+FP+FN)}$ is the percentage of correctly classified samples.
    \item
    In the effort-aware theme of this paper, we are   interested in the labelling effort to commission new models building which is \textbf{Cost} = $ \frac{|\{\text{human verified comments}\}|}{|\{\text{comments}\}|}$. 
    \item Except for FAR and Cost, for the rest of these metrics (Accuracy, Recall, and AUC), the {\underline{\em higher}} the value,
the 
{\underline{\em better}} the performance.
    
\end{itemize}
  {\bf Cautionary note:}   \citet{Menzies:2007prec} warns that precision can be misleading for imbalanced data sets like that studied here
(e.g.   Table~\ref{table:distribution} reports that for static warning analysis, the median of target class is 15\%).  Hence, while   we do not place much weight on
classifiers that fail on precision or F1.

\subsubsection{Statistical Analysis}

With the deterministic nature, we employed Cohen'$d$ effect size test to determine which results are similar by calculating $medium\_{step2}$ across Recall, False Alarm, AUC, Accuracy, and cost. As to what $d$ to use for this analysis, we take the advice of a widely accepted Sawilowsky et al.'s work~\cite{Sawilowsky2009NewES}. That paper asserts that ``small'' and ``medium'' effects can be measured using $d = 0.2$ and $d = 0.5$ (respectively). Splitting the difference, we will analyze this data looking for differences larger than $d = (0.5 + 0.2)/2 = 0.35$: \vspace{-10pt}


\begin{equation}
\small
\label{eq:cohen_step2}
Medium_{step2} \text{ or } M = 0.35 \cdot StdDev(\text{All results})
\end{equation}
The SOTA adoptable code warnings identifier and the SOTA issue close time predictor also validated their results with this test but with $d=0.35$ and $d=0.3$ respectively. 

\section{Labelling} \label{sec:motivations}

One of the goals of industrial analytics is that new conclusions can be quickly obtained from new data just by applying data mining
algorithms. As shown in Figure \ref{fig:pipeline}, there are at least nine separate
stages that must be completed before that goal be reached~\cite{amershi_ICSE19AI4SE}. Each of these stages offers
unique and separate challenges, each of which deserves extensive attention. Many of these steps have been extensively studied in
the literature \cite{liu2018satd,jitterbug,potdar2014exploratory,de2015contextualized,de2016investigating,de2020identifying,maldonado2015detecting,maldonado2017using,liu2018satd,huang2018identifying,zampetti2019automatically}. However, the labelling work of  step 4  has been receiving scant attention. In literature, there are several approaches for executing the labelling process:
\be
\item Manual labelling;
\item Crowdsourcing;
\item Reuse of labels;
\item Automatic labelling;
\item Active learning (a special kind of semi-supervised learning)
\ee

All of these approaches have their drawbacks; e.g. they are error-prone or will not scale. In response to these shortcomings, this study will take two directions:
\bi
\item First, we will try a {\em label-free} approach using  a plain  {\em unsupervised learning} technique to label the data;

\item If the {\em label-free} approach fails, then we  try  {\em tuned semi-supervised learning}, called FRUGAL, which optimizes the unsupervised learner's configurations in the grid search manner while validating the results on only 
\revised{2.5\%} of the labelled data. 
\ei

\subsection{Manual Labelling}

In  manual labelling,
a team of (e.g.)   graduate students assigns labels then     (a) cross-checks their work
 via   say, a Kappa statistic; then (b)~use some  skilled third person to resolve any labelling disagreements~\cite{lutz2004empirical,tu2020changing,tu2020better}.
 
Manual labelling   can be very slow. Tu et al. recently studies a corpus of 678 Github projects~\cite{tu2020better,tu2020changing}.
A  random selection of 10 projects from that corpus had
   $22,500$  commits, which took 175 hours to manually label the commits {\em buggy, non-buggy} (time includes cross-checking). That is,  manual labelling
   of  those   500 projects would  have
   required 90 weeks of   work.
   
   \subsection{Crowdsourcing}
   Tu et al. \cite{tu2020better} offers a cost estimate of what resources would be required to sub-contract that effort to   dozens of  crowdsourced workers via tools like  Mechanical Turk (MT). Applying best practices in crowdsourcing~\cite{chen2019replication}, assuming (a)~at least USA minimum \revised{wages} ~\cite{silberman2018responsible};
   and (b)~our university taking a 50\% overhead tax on grants; then   crowd sourcing the labelling  of the issues from 
500 projects  would require   \$320,000 of grant reserve.

   \subsection{Reusing Labels}
   
 Since manual labelling is time consuming and crowdsourcing is too expensive, researchers often   reuse   labels from previous studies.  E.g. for defect prediction, researchers \cite{xia16ist17,yang16unsupervised,yang2015deep} certified  their methods  using data generated by Kamei et al. \cite{kamei12_jit}. 
 This approach fails,
 in two cases.
 Firstly,
 when exploring a new domain, there may be no
 relevant old labels to reuse. 
 Secondly, reusing labels   means
 incorrectly labelled
 examples can 
 contamiante other research.
For example,     Yu et al.~\cite{jitterbug}   were exploring self-admitted technical debt and  found that their classifiers had an alarming high false-positive rate. But when they manually checked the labels of their data (which they taken from a prior study by Maldonado et al.~\cite{maldonado2015detecting}), they found  that over 98\% of the reused false-positive labels were incorrect. 
  
\subsection{Automatic Labelling} 

If labels cannot be generated manually or reused from other papers, using automatic labelling processes is an attractive alternative. For example,  defect prediction papers~\cite{commitguru,Kim08changes,catolino17_jitmobile,nayrolles18_clever,mockus00changeskeys,kamei12_jit,hindle08_largecommits}  can  label a commit as ``bug-fixing''
 when the commit text contains certain  keywords (e.g.
"bug", "fix", "wrong", "error", "fail" etc~\cite{tu2020better}).
 Vasilescu et al. \cite{Vasilescu15github}  noted that these keywords are used in a somewhat ad hoc manner
(researchers peek at a few results, then tinker   with  regular expressions that combine these keywords). 
Tu et al. \cite{tu2020better} had found that these
simplistic keyword approaches can introduce many errors,
perhaps due to the specialization of the project nature or the ad-hoc nature of their creation. In technical debts identification, Yu et al. proposed a pattern-based method that automatically identified 20-90\% of \revised{self-admitted technical debts (SATDs)} by finding patterns associated with high precision from the labelled training sets (close to 100\%). This  approach does need
extensively labelled training data  to find quality patterns that are associated with technical debt because it relies on precision. Another automatic approach is ML which involves supervised learning models to train on existing labelled datasets to learn the underlying rules of the data. However, this also requires having access to a substantial amount of labelled data (especially for deep learners) which is not always available in new domains (e.g., the success of open-source projects).

\vspace{-10pt}

\subsection{Active Learning}\label{actlearn}

A third approach is to (a)~only label a representative sample of the data; then (b)~build a classifier from that sample; then (c)~use that classifier to label the remaining data~\cite{ref49}.  To find that representative
example, some unsupervised learners like an associations rule learner or a clustering algorithm or an instance selection algorithm is used to find repeated patterns in the data~\cite{Kim08changes}. Then a human oracle is asked to label just one exemplar from each pattern. 
More sophisticated versions of this scheme include 
{\em active learners},
where an AI tool rushes ahead of the human to fetch the most information next examples to be labelled~\cite{me13a,settles2009active}.  
If humans first label most informative examples, then better models can be built faster. This means,
in turn, that humans have to label
fewer examples.

The more general term for  {\em active learning}  is  {\em semi-supervised learning}. Both terms mean ``do what you can with a small sample of the labels'' while {\em active learning} adds   a feedback loop   that checks  new
labels, one at a time, with some oracle. Semi-supervised learning relies on partially labelled data and mostly unlabelled data. 


Since 2012, active learning approaches have been received scarce attention in SE~\cite{kocaguneli2012active,tu2020better,yu2019improving,yu2020identifying}.
Initially, it seems to be   a   promising method for addressing the cost of label checking and generating.
 For self-admitted technical debt identification, only 24\% on the median of the  training corpus had to be labelled~\cite{yu2020identifying};
Also,  using active learning,
effort estimation for $N$ projects only needed labels on    11\% of those projects~\cite{kocaguneli2012active};
Further, 
while seeking 95\% of the vulnerabilities in 28,750  Mozilla Firefox C and C++ source code files,   humans only had to  inspect 30\% of the code~\cite{yu2019improving}.
That said,  after much work, it must be reported that 
 active learning still produces disappointing results.
 It is still   daunting   to
``only''  label (say)  5\% to 2.5\% of the projects in the   1,857,423 projects in RepoReapers~\cite{curating} or
the 
9.6 million links  explored by Hata et al.~\cite{Hideaki19}.
Also, consider the Firefox study mentioned in the last paragraph. The human effort of inspecting 28,750×30\% = 8,625 source code files 
(needed to find identify 95\% of the vulnerabilities)
it is beyond the resources of most analysts
(but it
might be justified for  mission-critical projects). \revised{Finally, for defect prediction, \citet{nsglp} proposed NSGLP and certified their method by varying the size of labeled software modules from 10 to 30\% of all the NASA datasets. The differences between our approach and their are listed in the Table~\ref{tbl:diffs}. They claimed that the proposed method outperformed several representative state-of-the-art semi-supervised ones for software defect prediction. However, reproducing that paper is complex since
it  was written before the current focus on research paper artifacts 
(so that paper has no reproduction package). Moreover, recent and widely cited studies 
argue  that the datasets used in that analysis  are
of dubious quality~\cite{data_jinx, quality}.}

\begin{table}[!t]
\vspace{-10pt}
\caption{Differences between FRUGAL and ~\citet{nsglp}}
\vspace{-10pt}
\label{tbl:diffs}
\footnotesize
\begin{center}
\begin{tabular}{l|p{2.58cm}|p{2.5cm}}
& \textbf{FRUGAL (this paper)}   & \textbf{~\citet{nsglp}} \\ \hline
\multirow{1}{*}{\rotatebox[origin=c]{0}{\parbox[c]{1cm}{ \textbf{Core}  \\ \textbf{Assumption}}}}   & Applies SE domain knowledge; i.e. higher complexity is associated with target instances (which we measure as being above $C$) & 
Applies graph theory; i.e. continuity and clustering (similar things have similar properties).      \\ \hline
\textbf{Hyperparameter} & Yes & No \\
\textbf{Optimization} & & \\\hline
\multirow{1}{*}{\rotatebox[origin=c]{0}{\parbox[c]{1cm}{ \textbf{Class}  \\ \textbf{Rebalancing}}}} & No  & Yes (with the Laplacian score sampling strategy) \\
\end{tabular}
\end{center}
\vspace{-5pt}
\end{table}

\begin{algorithm}[!tb]
\caption{Pseudocode of FRUGAL}
\footnotesize
\DontPrintSemicolon
    \KwInput{train\_data, tune\_data, test\_data}
    \KwOutput{result}
    
    percentiles = range(5, 100, 5) \;
    methods = [CLA, CLA\_ML, CLAFI\_ML]  \;
    best\_result = -1 \;
    best\_model = None \;
    \For{M in methods}
    {
        \For{C in percentiles}
        {
            model = M(C).fit(train\_data) \;
            temp\_result = model.predict(tune\_data) \;
            \If{isBetter(temp\_result, best\_result)}
            {
                best\_result = temp\_result \;
                best\_model = model \;
            }
        }
    }
    \Return{best\_model.predict(test\_data)}

\label{algo:frugal}
\end{algorithm}

It is opportune that in this adoptable static code warnings identification and issue close time prediction work, we aim to reduce the reviewing cost of these labelling methods by two methods (1) label-free approach with unsupervised learning, or (2) tuned semi-supervised learning to optimize the unsupervised learner's configurations in the grid search manner while validating the results on a small amount of the labelled data, i.e., \revised{2.5\%}.

\section{Methodology} \label{sec:methodology}

\subsection{General Framework}

Our approach, \revised{shown in Algorithm~1},
extends unsupervised learning with some semi-supervised learning and tuning. Nam et al.'s CLA is the SOTA unsupervised learner for defect prediction, which is also confirmed by \citet{unsup_review}'s large-scale study. As shown in Figure~\ref{fig:cla}, CLA consists of three modes: CLA, CLA+ML, and CLAFI+ML. This study shall adopt and extend CLA with tuning in the grid search manner of (1) three modes of CLA while varying (2) the $C\%$ percentile parameter. Simply, \revised{as illustrated in Algorithm 1}, FRUGAL finds the best combination of unsupervised learners = \{CLA, CLA+ML, CLAFI+ML\} and $C = \{5\%$ to $95\%$ increments by $5\%\}$. The author only proposed CLA and CLAFI+ML but CLA+ML is a natural medium that can be useful during the tuning process. 

We   explain the details of our approach in   \S4.2, \S4.3, and \S4.4.

\subsection{CLA} \label{sec:cla}

In the SOTA comparative study of unsupervised models in defect prediction, 
CLA starts with two steps of (1) \underline{\textbf{C}}lustering the instances and (2) \underline{\textbf{LA}}belling those instances accordingly to the cluster. In the setting with no train data available, we can label or predict all new/test instances, as shown in the \colorbox[HTML]{f0f9e8}{first block} of Figure~\ref{fig:cla}. 

\noindent \underline{\textbf{Clustering}}: 
\be
\item Find the median of feature $F_1, F_2,..., F_n$ ($percentile(F_i, C)$) where $C = 50\%$ across the whole dataset.
\item For each data instance $X_i$, go through each feature value of the respective data instance to count the time when the feature $F_i > percentile(F_i, C)$ as $K_i$.
\ee

\noindent \underline{\textbf{Labelling}}: label the instance $X_i$ as the positive class if $K_i > median(K)$, else label it as the negative class.

The intuition of such methods is based on the defect proneness tendency that is often found in defect prediction research, that is \textit{the higher complexity is associated with the proneness of the defects}  \cite{nam2015clami}. Simply, there is a tendency where the problematic instance's feature values are higher than the non-problematic ones. This tendency and CLA's/CLAFI+ML's effectivenesses are confirmed via the recent literature and comparative study of 40 unsupervised models in defect prediction across 27 datasets and three types of features by \citet{unsup_review}. They found CLA's/CLAFI+ML's performances are superior to other unsupervised methods while similar to supervised learning approaches. Therefore, this study investigated and found that the hypothesized tendency is also applicable in issues close time prediction and adoptable static code warning identification data but not with $C$ at the median ($C = 50\%$). This opens opportunities for hyperparameter tuning. 

\subsection{CLA + ML} \label{sec:cla+ml}

If there is an abundant train data in the wild but without labels, CLA can pseudo-label the train data before applying any machine learner in the ``supervised'' manner (as shown in the \colorbox[HTML]{bae4bc}{second block} in Figure~\ref{fig:cla}). For this step, we take \citet{nam2015clami}'s advice to incorporate Random Forest~\cite{Breiman2001} (RF, described in \S4.5.1), an ensemble of tree learners method, as the machine learner of choice.

\subsection{CLAFI + ML} \label{sec:clafi+ml}

CLAFI is an extension of CLA which is a fullstack framework that also include (3) \underline{\textbf{F}}eatures selection and (4) \underline{\textbf{I}}nstances selection. The setting is similar to CLA+ML, as shown in the \colorbox[HTML]{7bccc4}{third block} of Figure~\ref{fig:cla}, the pseudo-labelled train data (from CLA) and unlabelled test data will be processed with \underline{\textbf{FI}} and \underline{\textbf{F}} respectively. Finally, machine learner can train the processed pseudo-labelled train data and then predict on the processed test data.   


\noindent \textbf{\underline{Feature Selection}}: Calculate the violation score per feature, called metric in the original proposal of Nam et al. \cite{nam2015clami}. The process is done on both the train and the test dataset. 
\be
\item For each $F_i$, go through all instances of $X_j$, a violation happens when $F_i$ at $X_j$ is higher than the $percentile(K_i, C)$ where $C = 50\%$ but $Y_j=1$ and vice-versa. 
\item Sum all the violations per feature across the whole dataset and sort it in ascending order.
\item Select the feature with the lowest violation score, if multiple of them have the same score then pick all of them. 
\ee 

\noindent \textbf{\underline{Instance Selection}}: 
\be
\item With the selected features, go through each instance $X_i$ and check if the respective $F_j$ values violated the proneness assumption then remove that instance $X_i$.  
\item If the dataset do not have instances with both classes at the end then pick the next minimum violation score to select metrics. 
\item This process is only done on the train dataset.
\ee

After selecting features with the minimum violation scores and removing the instances that violated the proneness tendency, a practitioner can train an RF model on the processed train data to identify the target classes from the processed test dataset. 





\subsection{Machine Learning Models}

\vspace{5pt}\subsubsection{Random Forest (RF)}  is an ensemble learning method that operates by constructing a multitude of decision trees,
each time with different subsets of the data rows $R$ and columns $C$\footnote{Specifically, using $\log_2{C}$ of the columns, selected at random.}.
Each decision tree is recursively built to
find the features that reduce most of {\em entropy}, where a higher entropy indicates less ability to draw conclusions from the data being processed~\cite{CART}.
Test data is then passed across all $N$ trees and the conclusions are determined (say) a majority vote across all the trees~\cite{Breiman2001}. Holistically, RF is based on bagging (bootstrap aggregation) which averages the results over many decision trees from sub-samples (reducing variance).

\subsubsection{Support Vector Machine (SVM)}  is a  classifier  defined by a separating hyperplane~\cite{suykens1999least}. Soft-margin linear SVMs are commonly used in text classification given the high dimensionality of the feature space. This was recommended by \citet{intrinsic_static} as the state of the art for our adoptable static code warning identification domain. A soft-margin linear SVM looks for the decision hyperplane that maximizes the margin between training data of two classes while minimizing the training error (hinge loss): \vspace{-10pt}
\begin{equation}
\label{eq:SVM}
\min {\lambda \lVert w\rVert ^{2}+\left[{\frac {1}{n}}\sum _{i=1}^{n}\max \left(0,1-y_{i}(w\cdot x_{i}-b)\right)\right]}
\end{equation}
where the class of $x$ is predicted as $sgn(w\cdot x-b)$.

Both SVM and RF are popular in the field of ML and implemented in the popular open-source toolkit Scikit-learn by \cite{pedregosa2011scikit}. 

\subsubsection{Feedforward Neural Networks} is the first and simplest technology devised from artificial neural network~\cite{SCHMIDHUBER201585}. The information moves in the forward direction only, starting from the input nodes through the hidden nodes and to the output nodes. At each node of these networks, the inputs are multiplied with weights that are learned, and then an activation function is applied. The weights are learned by the backpropagation algorithm \cite{rumelhart1985learning}. This uses just a few layers while the ``deep" learners use many layers. Also, the older
 methods use a threshold function at each node, while feedforward networks typically use the Rectified Linear Unit function \cite{nair2010rectified} of $f(x) = \max (0, x)$. This is the base learner for our second domain's state of the art where \citet{simple_ict} proposed a framework combining different preprocessors and different configurations of the simple feedforward neural network. 


\section{Results} \label{tion:result}

In order to make sure our proposed method's effectiveness is not affected by the bias between deterministic and non-deterministic models or the bias of uncertainty, we
randomly shuffle     train/test sets and incorporate stratified sampling with five bins (ensuring that the class distribution of the whole
data is replicated in each bin). The process is repeated for the train data but also includes an extra \revised{2.5\%} validating partition  for each \revised{97.5\%} tuning partition. \revised{The median 2.5\% of labelled train data for static warning analysis and issue close time are 36 and 1120 respectively}. During the simulation, the tune partition will not review labels for our unsupervised learning and semi-supervised learning candidates. FRUGAL  does have access to the corresponding \revised{2.5\%} labelled validation partition while deciding on the best configurations. For each 20\% of the test data, the process learns a model on five stratified samples of the train data. \revised{This process is done for both domains in this paper.} \\

\noindent \textbf{RQ1: How much labelled data ($L$\%) that FRUGAL requires?}

\begin{figure}[!b]
  \begin{subfigure}[t]{0.48\linewidth}
    \centering
    AUC in Actionable Static Warning identification.
    ``medium effect'' or $M$ = 5\% \vspace{10pt}
    \includegraphics[width=1\linewidth]{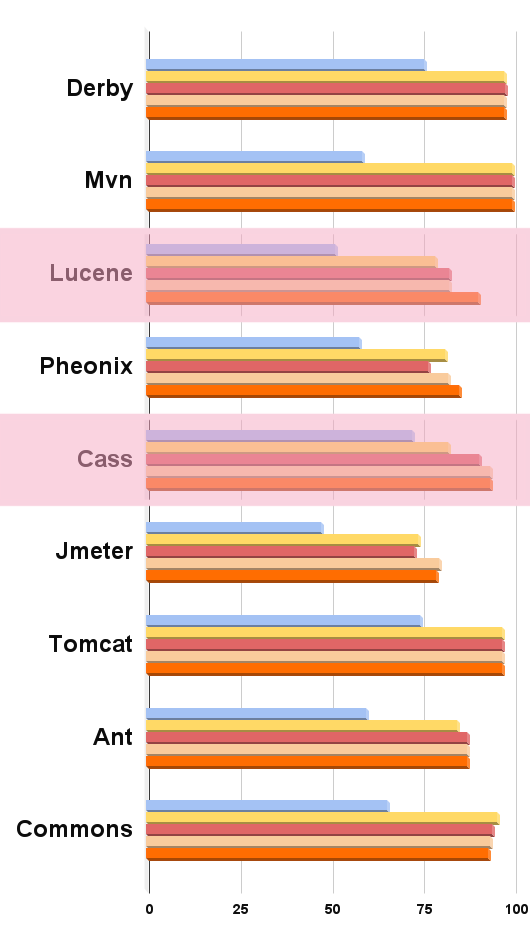}

  \end{subfigure}
  \hfill
  \begin{subfigure}[t]{0.48\linewidth}
  \centering
    Accuracy in Issues Close Time prediction. \\
    ``medium effect'' or $M$ = 2\% \vspace{10pt}
    \includegraphics[width=1\linewidth]{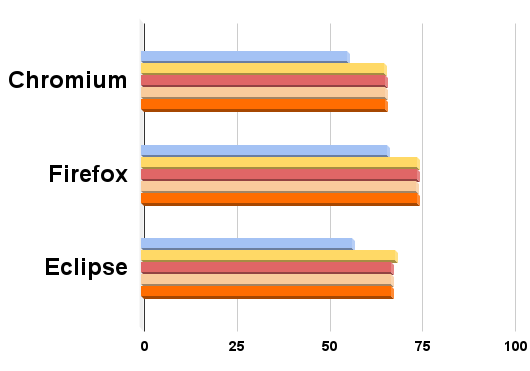}
  \end{subfigure}
  \begin{subfigure}[b]{\linewidth}
    \footnotesize
    \centering
    \includegraphics[width=0.65\linewidth]{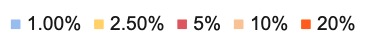}
  \end{subfigure}
  \vspace{-20pt}
  \caption{RQ1 results on two \revised{domains} with FRUGAL($L \in \{1\%, 2.5\%, 5\%, 10\%, 20\%\}$). \revised{FRUGAL with $L > 1\%$ perform similarly. However, for adoptable static warning identification, FRUGAL($L=2.5\%$) does perform the best across 7 datasets except in Lucene and Cass (the \colorbox[HTML]{FFCCC9}{highlighted} ones) where  FRUGAL(L=$20\%$) outperforms FRUGAL(L=$2.5\%$).}}
  \label{fig:rq1} 
\end{figure}

Our hypothesis is ``there are few key data regions where extra data would lead to indistinguishable results''. We test the different amounts of the train data's labels that are required for FRUGAL's performance to plateaus. Let $L$ be 1\%, 2.5\%, 5\%, 2.5\%, or 20\%, Figure \ref{fig:rq1} reports FRUGAL's performance on both adoptable static warning identification (in AUC) and issues close time prediction (in accuracy). 
Both metrics are derived from the SOTA's evaluation metrics. Specifically, we pick AUC as the representation metric for adoptable static warning analysis since AUC measures the area under the curve whereas other metrics only calculate a single point on the curve.  From Figure \ref{fig:rq1}:
 \bi
\item \revised{The lower bound for both domains is FRUGAL with $L = 1\%$.}
\item \revised{For adoptable static warning identification, FRUGAL's performance improves initially and  plateaus beyond $L = 2.5\%$ across 7 datasets. FRUGAL($L=2.5\%$) loses to FRUGAL($L=20\%$) in only \textit{Lucene} and \textit{Cass} projects but reduces 8 (20\%/2.5\%) times the labelling efforts.}
\item For issues close time prediction, FRUGAL surprisingly performs statistically similar across all $L \in \{2.5\%, 5\%, 2.5\%, 20\%\}$. 
\ei

The same effect is absent in issue close time prediction, this is highly likely due to the balanced nature of the data's class distribution. However, the data in static warning analysis is more imbalanced (with a median of 15\% for the adoptable static warning class ratio). This is consistent with the motivations for oversampling and undersampling techniques for imbalanced data~\cite{chawla2002smote, agrawal2018better}. \\

\begin{RQ}{\normalsize{In summary, our answer to RQ1 is:}} 
From our investigation of various $L$ values, FRUGAL's performance plateaus beyond \revised{$L\geq2.5\%$} and FRUGAL's success is not altered by large changes to $L$.  
\end{RQ}

\vspace{5pt}

\begin{table*}[!t]
\footnotesize
\caption{Comparison between \textbf{CLA}\cite{nam2015clami}, \textbf{SVM}\cite{intrinsic_static}, and FRUGAL in terms of FAR, Recall, Precision, F1, and AUC for identifying adoptable static warning.
In this table, the FRUGAL results were  found after labelling just  2.5\% of the data.
Except for FAR, the higher the results the better the performance of the treatment. Medians and IQRs (delta between 75th
and 25th percentile, lower the better) are calculated for easy comparisons. Here, the \colorbox[HTML]{FFCCC9}{highlighted} cells show best performing treatments.}
\centering
\setlength\tabcolsep{5pt}
\begin{tabular}{c|l|c|c|c|c|c|c|c|c|c|c|c}
   \textbf{Metrics} & \textbf{Treatment}  & \textbf{\rotatebox[origin=c]{90}{Derby}} & \textbf{\rotatebox[origin=c]{90}{Mvn}} & \textbf{\rotatebox[origin=c]{90}{Lucene}} & \textbf{\rotatebox[origin=c]{90}{Phoenix}} & \textbf{\rotatebox[origin=c]{90}{Cass}} & \textbf{\rotatebox[origin=c]{90}{Jmeter}} & \textbf{\rotatebox[origin=c]{90}{Tomcat}} 
   & \textbf{\rotatebox[origin=c]{90}{Ant}}
   & \textbf{\rotatebox[origin=c]{90}{Commons}} & \textbf{\rotatebox[origin=c]{90}{Median}} & \textbf{\rotatebox[origin=c]{90}{IQR}} \vspace{2pt}\\

\hline
& 	CLA		& 	43.3	& 	44.9	& 	30.8	& 	39.1	& 	42.1	& 	39.4	& 	43	& 	45.8	& 	43	& 	43	& 	4.2 \\ 
FAR & 	CLAFI+RF		& 	\cellcolor[HTML]{FFCCC9}0.4	& 18.1	& \cellcolor[HTML]{FFCCC9}0.8	& 29.4	& \cellcolor[HTML]{FFCCC9}0.5	& \cellcolor[HTML]{FFCCC9}4.1	& \cellcolor[HTML]{FFCCC9}0.4	& 22.5	& 13.6	& 4.1	& 18.5 \\ 
	\rowstyle{\color{black}}($M = 6\%$)& 	FRUGAL	& 	\revised{\cellcolor[HTML]{FFCCC9}1}	& 	\revised{\cellcolor[HTML]{FFCCC9}4.8}	& 		\revised{\cellcolor[HTML]{FFCCC9}3.1}	& 		\revised{\cellcolor[HTML]{FFCCC9}2.1}	& 		\revised{\cellcolor[HTML]{FFCCC9}0.6}	& 		\revised{\cellcolor[HTML]{FFCCC9}8.0}	& 		\revised{\cellcolor[HTML]{FFCCC9}2.8}		& 	\revised{\cellcolor[HTML]{FFCCC9}3.7}	& 		\revised{\cellcolor[HTML]{FFCCC9}7.8} & \revised{3.1} &	\revised{2.9}\\ 
	& \rowstyle{\color{black}}SVM~\cite{intrinsic_static} (L=2.5\%)	&\rowstyle{\color{black}}	\cellcolor[HTML]{FFCCC9}0.7	&\rowstyle{\color{black}}	100	&\rowstyle{\color{black}}	\cellcolor[HTML]{FFCCC9}5.1	&\rowstyle{\color{black}}	\cellcolor[HTML]{FFCCC9}0.5	&\rowstyle{\color{black}}	\cellcolor[HTML]{FFCCC9}0.8	&\rowstyle{\color{black}}	10.5	&\rowstyle{\color{black}}	\cellcolor[HTML]{FFCCC9}2.1	&\rowstyle{\color{black}}	\cellcolor[HTML]{FFCCC9}0.2	&\rowstyle{\color{black}}	100	&\rowstyle{\color{black}}	2.1	&\rowstyle{\color{black}}	32.2 \\
	& 	SVM~\cite{intrinsic_static}	& 	\cellcolor[HTML]{FFCCC9}1.3	& 	\cellcolor[HTML]{FFCCC9}1.2	& 	6.9	& 	\cellcolor[HTML]{FFCCC9}3.5	& 	\cellcolor[HTML]{FFCCC9}1.4	& 	\cellcolor[HTML]{FFCCC9}2.1	& 	\cellcolor[HTML]{FFCCC9}3.2	& 	\cellcolor[HTML]{FFCCC9}0.5	& 	\cellcolor[HTML]{FFCCC9}5.8	& 	2.1	& 	2.7 \\

\hline
	& 	CLA	& 	45.8	& 	\cellcolor[HTML]{FFCCC9}100	& 	64.5	& 	66.7	& 	\cellcolor[HTML]{FFCCC9}98.6	& 	75.9	& 	63.1	& 	80	& 	\cellcolor[HTML]{FFCCC9}100	& 	75.9	& 	32.8 \\
	Recall & 	CLAFI+RF		& 	57.2	& \cellcolor[HTML]{FFCCC9}93.3	& 67.1	& 62.1	& 83.7	& 73.1	& 64.4	& 77.8	& 77.5	& 73.1	& 12.9 \\ 
	 \rowstyle{\color{black}} ($M = 9\%$) & 	FRUGAL	&  		\revised{\cellcolor[HTML]{FFCCC9}93.7}	& 		\revised{\cellcolor[HTML]{FFCCC9}92.9}	& 		\revised{\cellcolor[HTML]{FFCCC9}100}	& 	 		\revised{\cellcolor[HTML]{FFCCC9}98.7}	& 		\revised{\cellcolor[HTML]{FFCCC9}92.6}	& 		\revised{\cellcolor[HTML]{FFCCC9}88.4}	& 		\revised{\cellcolor[HTML]{FFCCC9}98.8}	& 		\revised{\cellcolor[HTML]{FFCCC9}94.2}	& 		\revised{\cellcolor[HTML]{FFCCC9}99}	& 		\revised{94.2}	& 		\revised{6} \\
	 & \rowstyle{\color{black}}SVM~\cite{intrinsic_static} (L=2.5\%) &\rowstyle{\color{black}} 0	&\rowstyle{\color{black}}0	&\rowstyle{\color{black}}73.6	&\rowstyle{\color{black}}65.2	&\rowstyle{\color{black}}31.9	&\rowstyle{\color{black}}31.4	&\rowstyle{\color{black}}78.9	&\rowstyle{\color{black}}58.5	&\rowstyle{\color{black}}0 &\rowstyle{\color{black}} 31.9&\rowstyle{\color{black}}	43.9 \\
	& 	SVM~\cite{intrinsic_static}	& 	\cellcolor[HTML]{FFCCC9}97.8	& 	\cellcolor[HTML]{FFCCC9}97	& 	87.1	& 	\cellcolor[HTML]{FFCCC9}96.1	& 	\cellcolor[HTML]{FFCCC9}90.3	& 	\cellcolor[HTML]{FFCCC9}93.3	& 	\cellcolor[HTML]{FFCCC9}98.2	& 	\cellcolor[HTML]{FFCCC9}95	& 	\cellcolor[HTML]{FFCCC9}99.5	& 	96.1	& 	4.8 \\

\hline
	 & 	\rowstyle{\color{black}}CLA	& 	\rowstyle{\color{black}}4.8	&	\rowstyle{\color{black}}6.6	&	\rowstyle{\color{black}}49.6	&	\rowstyle{\color{black}}22.3	&	\rowstyle{\color{black}}27.3	&	\rowstyle{\color{black}}38.3	&	\rowstyle{\color{black}}31.9	&	\rowstyle{\color{black}}8.2	&	\rowstyle{\color{black}}11.1	&	\rowstyle{\color{black}}22.3	&	\rowstyle{\color{black}}23.1 \\ 
	\rowstyle{\color{black}}Precision & 	\rowstyle{\color{black}}CLAFI+RF 	& \rowstyle{\color{black}}1.9	&	\rowstyle{\color{black}}4.3	&	\rowstyle{\color{black}}39.3	&	\rowstyle{\color{black}}15.5	&	\rowstyle{\color{black}}24.1	&	\rowstyle{\color{black}}22	&	\rowstyle{\color{black}}18	&	\rowstyle{\color{black}}4.6	&	\rowstyle{\color{black}}9	&	\rowstyle{\color{black}}15.5	&	\rowstyle{\color{black}}14.6 \\ 
	\rowstyle{\color{black}}($M = 17\%$) & 	\rowstyle{\color{black}}FRUGAL	& 	\rowstyle{\color{black}}61.4	&	\rowstyle{\color{black}}63.5	&	\rowstyle{\color{black}}60.9	&	\rowstyle{\color{black}}61.3	&	\rowstyle{\color{black}}67	&	\rowstyle{\color{black}}50.6	&	\rowstyle{\color{black}}67.9	&	\rowstyle{\color{black}}32.6	&	\rowstyle{\color{black}}\cellcolor[HTML]{FFCCC9}78.4	&	\rowstyle{\color{black}}61.4	&	\rowstyle{\color{black}}8.6 \\
	& \rowstyle{\color{black}}SVM~\cite{intrinsic_static} (L=2.5\%) 	&	\rowstyle{\color{black}} 0 	&	\rowstyle{\color{black}}	0 	&	\rowstyle{\color{black}}	\cellcolor[HTML]{FFCCC9}86.4 	&	\rowstyle{\color{black}}	\cellcolor[HTML]{FFCCC9}85 	&	\rowstyle{\color{black}}	\cellcolor[HTML]{FFCCC9}90.8		&	\rowstyle{\color{black}}54		&	\rowstyle{\color{black}}\cellcolor[HTML]{FFCCC9}91.8		&	\rowstyle{\color{black}}\cellcolor[HTML]{FFCCC9}94		&	\rowstyle{\color{black}}0		&	\rowstyle{\color{black}}85		&	\rowstyle{\color{black}}55.6 \\
	 & 	\rowstyle{\color{black}}SVM~\cite{intrinsic_static}	& 	\rowstyle{\color{black}}\cellcolor[HTML]{FFCCC9} 86.4	& 	\rowstyle{\color{black}}\cellcolor[HTML]{FFCCC9}100	& 	\rowstyle{\color{black}}\cellcolor[HTML]{FFCCC9}72.7	& 	\rowstyle{\color{black}}\cellcolor[HTML]{FFCCC9}84.3	& 	\rowstyle{\color{black}}\cellcolor[HTML]{FFCCC9}100	& 	\rowstyle{\color{black}}\cellcolor[HTML]{FFCCC9}71.4	& 	\rowstyle{\color{black}}\cellcolor[HTML]{FFCCC9}83.1	& 	\rowstyle{\color{black}}\cellcolor[HTML]{FFCCC9}90	& 	\rowstyle{\color{black}} 33.3	& 	\rowstyle{\color{black}}84.3 & 	\rowstyle{\color{black}}	20 \\

 \hline
	 & 	\rowstyle{\color{black}}CLA	& 	\rowstyle{\color{black}}8.7	&	\rowstyle{\color{black}}12.3	&	\rowstyle{\color{black}}54.2	&	\rowstyle{\color{black}}33.5	&	\rowstyle{\color{black}}42.9	&	\rowstyle{\color{black}}51.7	&	\rowstyle{\color{black}}42.7	&	\rowstyle{\color{black}}14.8	&	\rowstyle{\color{black}}20	&	\rowstyle{\color{black}}33.5	&	\rowstyle{\color{black}}26.4 \\
	\rowstyle{\color{black}}F1 & 	\rowstyle{\color{black}}CLAFI+RF		& 	\rowstyle{\color{black}}3.3	&	\rowstyle{\color{black}}8.1	&	\rowstyle{\color{black}}39.2	&	\rowstyle{\color{black}}26.5	&	\rowstyle{\color{black}}35.6	&	\rowstyle{\color{black}}27.3	&	\rowstyle{\color{black}}22.7	&	\rowstyle{\color{black}}8.4	&	\rowstyle{\color{black}}15.9	&	\rowstyle{\color{black}}22.7	&	\rowstyle{\color{black}}15.4 \\ 
	\rowstyle{\color{black}}($M = 16\%$) & 	\rowstyle{\color{black}}FRUGAL	& 	\rowstyle{\color{black}}63.6	&	\rowstyle{\color{black}}72.4	&	\rowstyle{\color{black}}\cellcolor[HTML]{FFCCC9}62.1	&	\rowstyle{\color{black}}45.6	&	\rowstyle{\color{black}}55.3	&	\rowstyle{\color{black}}45.4	&	\rowstyle{\color{black}}\cellcolor[HTML]{FFCCC9}69.1	&	\rowstyle{\color{black}}25.1	&	\rowstyle{\color{black}}\cellcolor[HTML]{FFCCC9}58.9 & \rowstyle{\color{black}}58.9  & 	\rowstyle{\color{black}}18.3 \\
	& \rowstyle{\color{black}}SVM~\cite{intrinsic_static} (L=2.5\%)	&\rowstyle{\color{black}}0	&\rowstyle{\color{black}}0	&\rowstyle{\color{black}}\cellcolor[HTML]{FFCCC9}66.4	&\rowstyle{\color{black}}34.3	&\rowstyle{\color{black}}27	&\rowstyle{\color{black}}22.8	&\rowstyle{\color{black}}\cellcolor[HTML]{FFCCC9}72.7	&\rowstyle{\color{black}}71.1	&\rowstyle{\color{black}}0  &\rowstyle{\color{black}}27	&\rowstyle{\color{black}} 50.5 \\
	& 	\rowstyle{\color{black}}SVM~\cite{intrinsic_static}	& 	\rowstyle{\color{black}}\cellcolor[HTML]{FFCCC9}  	86.4 & 	\rowstyle{\color{black}}\cellcolor[HTML]{FFCCC9} 	100 & 	\rowstyle{\color{black}}\cellcolor[HTML]{FFCCC9} 	72.7 & 	\rowstyle{\color{black}}\cellcolor[HTML]{FFCCC9} 	84.3 & 	\rowstyle{\color{black}}\cellcolor[HTML]{FFCCC9} 	100 & 	\rowstyle{\color{black}}\cellcolor[HTML]{FFCCC9} 	71.4 & 	\rowstyle{\color{black}}\cellcolor[HTML]{FFCCC9} 	83.1 & 	\rowstyle{\color{black}}\cellcolor[HTML]{FFCCC9} 	90 & \rowstyle{\color{black}} 	\cellcolor[HTML]{FFCCC9}72.8 & 	\rowstyle{\color{black}} 	84.3 & 	\rowstyle{\color{black}} 	20.1 \\

\hline
	& 	CLA	& 	54.9	& 	88.9	& 	66.8	& 	64.7	& 	83.8	& 	72	& 	69.7	& 	68.5	& 	81.1	& 	69.7	& 	13.7 \\ 
	AUC & 	CLAFI+RF 	& 66.8	& 89	& 77.3	& 69.4	& 78.4	& 78	& 75.2	& 75.9	& 84.3	& 77.3	& 4.2 \\ 
	\rowstyle{\color{black}}($M = 11\%$) & 	FRUGAL	& 	\rowstyle{\color{black}}\cellcolor[HTML]{FFCCC9}95.3	& 		\rowstyle{\color{black}}\cellcolor[HTML]{FFCCC9}94.6	& 		\rowstyle{\color{black}}82.5	& 		\rowstyle{\color{black}}78.6	& 		\rowstyle{\color{black}}\cellcolor[HTML]{FFCCC9}90.4		& 	 \rowstyle{\color{black}}74	& 	\rowstyle{\color{black}}\cellcolor[HTML]{FFCCC9}	97	& 		\rowstyle{\color{black}}\cellcolor[HTML]{FFCCC9}89.6	& 	\rowstyle{\color{black}}\cellcolor[HTML]{FFCCC9}	94.1 	& 		\rowstyle{\color{black}}90.4	& 		\rowstyle{\color{black}}12.7 \\
	& \rowstyle{\color{black}}SVM~\cite{intrinsic_static} (L=2.5\%)	&\rowstyle{\color{black}}77	&\rowstyle{\color{black}}0	&\rowstyle{\color{black}}\cellcolor[HTML]{FFCCC9}88.3	&\rowstyle{\color{black}}\cellcolor[HTML]{FFCCC9}89.7	&\rowstyle{\color{black}}87.1	&\rowstyle{\color{black}}65.6	&\rowstyle{\color{black}}\cellcolor[HTML]{FFCCC9}95.1	&\rowstyle{\color{black}}75.3	&\rowstyle{\color{black}}0  &\rowstyle{\color{black}}77	&\rowstyle{\color{black}}39.5 \\
	& 	SVM~\cite{intrinsic_static}	& 	\cellcolor[HTML]{FFCCC9}99.5	& 	\cellcolor[HTML]{FFCCC9}99.6	& \cellcolor[HTML]{FFCCC9}	97.3	& 	\cellcolor[HTML]{FFCCC9}98.8	& 	\cellcolor[HTML]{FFCCC9}99.7	& 	\cellcolor[HTML]{FFCCC9}98.8	& 	\cellcolor[HTML]{FFCCC9}99.6	& 	\cellcolor[HTML]{FFCCC9}99.7	& 	\cellcolor[HTML]{FFCCC9}99	& 	99.5	& 	0.8 \\

\hline
\end{tabular}
\label{tab:rq2}
\end{table*}


\textbf{RQ2: How does FRUGAL perform in adoptable static warnings identification?} 

\citet{wang2018there}  proposed the ``golden set'' features along with the ML study where they employed RF, Decision Tree, and SVM (with the RBF kernel) with median AUC performances at 70\%, 64\%,  and 50\%. Yang et al. \cite{intrinsic_static}  extensively investigated different deep learners (DNN, CNN, RF, Decision Tree, and SVM) that pushed Wang et al.'s results to new higher watermarks in the area with median AUC performances at 99.5\%, 95.9\%, and 99.5\% with almost 45\%, 55\%, and 100\% relative improvements to  the same learner choices as \citet{wang2018there}. The default parameters in Weka (used by \citet{wang2018there}) are different to 
those used in SciKit-Learn (used by \citet{intrinsic_static}). For instance, Wang et al.'s SVM used RBF kernel while \citet{intrinsic_static}'s SVM used linear kernel.

\citet{intrinsic_static} proposed the standard linear SVM as the SOTA's adoptable static warning identifier. Table \ref{tab:rq2} reports the comparison of our proposed method FRUGAL, the SOTA's SVM (with 100\% labelled data and 2.5\% labelled data), and the baseline unsupervised learners (CLA \& CLAFI+RF) across FAR, recall, precision, F1 and AUC. In those results:

\bi
\item Standard unsupervised learner CLA/CLAFI+RF performs the worst as their default behavior is  clustering based on the median of the data which may not apply for all the data and especially in the static warning analysis. However, CLA's recalls are almost 100\% in a few cases (\textit{Mvn, Cass,} and \textit{Commons}) and CLAFI+RF's FAR are almost 0\% in more than half cases (\textit{Derby, Lucene, Cass, Jmeter} and \textit{Tomcat}). This indicates promising areas for tuning configurations of unsupervised learners.

\item \revised{The SOTA work originally evaluated their method on FAR, recall, and AUC. With the same comparison, FRUGAL performs better than the SOTA's SVM as FRUGAL wins in Recall and FAR while losing in AUC. However, when considering precision and F1, FRUGAL underperforms.
Recalling our cautionary note
(from the end of \S\ref{measures}),
precision (and hence F1) can be misleading for data sets where the target class is 
rare ~\cite{Menzies:2007prec}
(e.g. as shown in Table~\ref{table:distribution}, the median of target class is 15\%).  Therefore, overall,
we say   FRUGAL performs similarly to the SOTA's SVM.}

\item \revised{The SOTA work that was trained on only 2.5\% labelled data (i.e., SVM with $L=2.5\%$)  underperforms both the SOTA work with 100\% labelled data and FRUGAL. This illustrates that FRUGAL's effectiveness is not due to random sampling of the data.}

\item In term of labelling efforts, CLA is label-free, FRUGAL costs \revised{2.5\%}, and Yang et al.'s method costs 100\% because FRUGAL and the SOTA require \revised{2.5\%} and 100\% of the data to be labelled.  

\ei 

\begin{RQ}{\normalsize{In summary, our answer to RQ2 is:}} 
FRUGAL improves significantly from standard unsupervised learner CLA with  \revised{2.5\%} of data labelled as a tradeoff while performing similarly to the SOTA with \revised{97.5\%} fewer information. A simple method that explores small regions of data does no worse than methods that extensively learn the whole space.
\end{RQ}

\begin{table}[!t]
\footnotesize
\vspace{-5pt}
\caption{Comparison between \textbf{CLA}\cite{nam2015clami}, \textbf{SIMPLE}\cite{simple_ict}, and FRUGAL in terms of Accuracy, FAR, Recall, and AUC for predicting issue close time. Note that in this table, the FRUGAL results were  obtained after labelling just  2.5\% of the data.
Except for FAR, the higher the results the better the performance of the treatment. Medians and IQRs (delta between 75th
and 25th percentile, lower the better) are calculated for easy comparisons. Here, the \colorbox[HTML]{FFCCC9}{highlighted} cells show best performing treatments.}
\vspace{-10pt}
\centering
\setlength\tabcolsep{5pt}
\renewcommand{\arraystretch}{1.1}

\begin{tabular}{c|l|c|c|c|c|c}
  \textbf{Metrics}  & \textbf{Treatment} & \textbf{\rotatebox[origin=c]{90}{Chromium}} & \textbf{\rotatebox[origin=c]{90}{Firefox}} & \textbf{\rotatebox[origin=c]{90}{Eclipse}} & \textbf{\rotatebox[origin=c]{90}{Median}} & \textbf{\rotatebox[origin=c]{90}{IQR}} \vspace{2pt}\\
\hline
	&	CLA	&	53.6	&	57.1	&	57.6	&	57.4	&	2 \\ 
	Accuracy & CLAFI+RF & 50.2	&		53.9	&		51.3	&		52.6	&		1.9 \\
	($M = 3\%$) &	\rowstyle{\color{black}}FRUGAL	&	\rowstyle{\color{black}}65.3	&	\rowstyle{\color{black}}\cellcolor[HTML]{FFCCC9}74.2	&	\rowstyle{\color{black}}\cellcolor[HTML]{FFCCC9}68.2	&	\rowstyle{\color{black}}68.2	&	\rowstyle{\color{black}}4.5 \\
	&	SIMPLE~\cite{simple_ict}	&	\cellcolor[HTML]{FFCCC9}70.3	&	68.3	&	\cellcolor[HTML]{FFCCC9}68.8	&	68.6	&	1 \\

\hline
	&	CLA		&	34.9		&	26.9		&	37.3		&	32.1		&	5.2 \\
	FAR &	CLAFI+RF		&	35.1	&		\cellcolor[HTML]{FFCCC9}3.1	&		32.9	&		18	&		16 \\
	($M = 5\%$) &	\rowstyle{\color{black}}FRUGAL		&	\rowstyle{\color{black}}\cellcolor[HTML]{FFCCC9}2.2	&		\rowstyle{\color{black}}\cellcolor[HTML]{FFCCC9}2		&	\rowstyle{\color{black}}\cellcolor[HTML]{FFCCC9}2		&	\rowstyle{\color{black}}2		&	\rowstyle{\color{black}}0.1 \\
	&	SIMPLE~\cite{simple_ict}		&	33.1		&	32.1		&	22.5		&	27.3		&	5.3  \\

\hline
		& CLA	&	54.9	&	88.9	&	66.8	&	77.9	&	17 \\ 
		Recall & CLAFI+RF	&	38.3	&		45.4	&		38.1	&		41.8	&		3.7 \\ 
	($M = 8\%$) &	\rowstyle{\color{black}}FRUGAL	&	\rowstyle{\color{black}}\cellcolor[HTML]{FFCCC9}99.9	&	\rowstyle{\color{black}}\cellcolor[HTML]{FFCCC9}97.9	&	\rowstyle{\color{black}}\cellcolor[HTML]{FFCCC9}97	&	\rowstyle{\color{black}}97.9	&	\rowstyle{\color{black}}1.5 \\
	&	SIMPLE~\cite{simple_ict}	&	71.7	&	74.1	&	54	&	64.1	&	10.1 \\

\hline
	&	\rowstyle{\color{black}}CLA	&	\rowstyle{\color{black}}61	&	\rowstyle{\color{black}}76.4	&	\rowstyle{\color{black}}62.9	&	\rowstyle{\color{black}}62.9	&	\rowstyle{\color{black}}7.7 \\
	\rowstyle{\color{black}}Precision & \rowstyle{\color{black}}CLAFI+RF & \rowstyle{\color{black}}57.4	&	\rowstyle{\color{black}}72.6	&	\rowstyle{\color{black}}57.9	&	\rowstyle{\color{black}}57.9	&	\rowstyle{\color{black}}7.6 \\
	\rowstyle{\color{black}}($M = 3\%$) &	\rowstyle{\color{black}}FRUGAL	&	\rowstyle{\color{black}}\cellcolor[HTML]{FFCCC9}69.1	&	\rowstyle{\color{black}}\cellcolor[HTML]{FFCCC9}79.9	&	\rowstyle{\color{black}}\cellcolor[HTML]{FFCCC9}69.2	&	\rowstyle{\color{black}}69.2	&	\rowstyle{\color{black}}5.4 \\
	&	\rowstyle{\color{black}}SIMPLE~\cite{simple_ict}	&	\rowstyle{\color{black}}63.5	&	\rowstyle{\color{black}}67	&	\rowstyle{\color{black}}\cellcolor[HTML]{FFCCC9}70.4	&	\rowstyle{\color{black}}67	&	\rowstyle{\color{black}}3.5 \\

\hline
		& \rowstyle{\color{black}}CLA	&	\rowstyle{\color{black}}51.3	&	\rowstyle{\color{black}}59.1	&	\rowstyle{\color{black}}57.7	&	\rowstyle{\color{black}}57.7	&	\rowstyle{\color{black}}3.9 \\ 
		\rowstyle{\color{black}}\rowstyle{\color{black}}F1 & \rowstyle{\color{black}}CLAFI+RF	&	\rowstyle{\color{black}}46	&	\rowstyle{\color{black}}55.9	&	\rowstyle{\color{black}}45.9	&	\rowstyle{\color{black}}46	&	\rowstyle{\color{black}}5  \\ 
	\rowstyle{\color{black}}($M = 4\%$) &	\rowstyle{\color{black}}FRUGAL	&	\rowstyle{\color{black}}\cellcolor[HTML]{FFCCC9}75	&	\rowstyle{\color{black}}\cellcolor[HTML]{FFCCC9}82.3	&	\rowstyle{\color{black}}\cellcolor[HTML]{FFCCC9}74.8	&	\rowstyle{\color{black}}75	&	\rowstyle{\color{black}}3.75 \\
	&	\rowstyle{\color{black}}SIMPLE~\cite{simple_ict}	&	\rowstyle{\color{black}}63.5	&	\rowstyle{\color{black}}61.4	&	\rowstyle{\color{black}}57.9	&	\rowstyle{\color{black}}61.4	&	\rowstyle{\color{black}}2.8 \\

\hline
	&	CLA	&	58.8	&	66.4	&	62.2	&	64.3	&	3.8 \\
	AUC & CLAFI+RF & 53.4	&		60.6	&		54.2	&		57.4	&		3.6 \\
	\rowstyle{\color{black}} ($M = 3\%$) & \rowstyle{\color{black}}	FRUGAL	&	\rowstyle{\color{black}}\cellcolor[HTML]{FFCCC9}72.1	&	\rowstyle{\color{black}}\cellcolor[HTML]{FFCCC9}80.2	&	\rowstyle{\color{black}}\cellcolor[HTML]{FFCCC9}75.8	&	\rowstyle{\color{black}}75.8	&	\rowstyle{\color{black}}4.1 \\
	&	SIMPLE~\cite{simple_ict}	&	67.3	&	70.4	&	65.6	&	68	&	2.4 \\

\hline
\end{tabular}
\label{tab:rq3}
\vspace{-15pt}
\end{table}

\vspace{5pt}

 \textbf{RQ3: How well
 does FRUGAL predict    issue close time?}

Mark et al.\cite{mani2019deeptriage} proposed DeepTriage as SOTA deep learning solution  extended from bidirectional LSTMs with an ``attention mechanism'' to predict issue close time. A Long Short-Term Memory (LSTM) \cite{hochreiter1997long} is a    recurrent neural network with an
additional ``gate" mechanisms to allow the network to model connections between long-distance tokens in the input. Bidirectional variants of recurrent models, such as LSTMs, use  token stream in both forward and backward directions;  allowing for the network to model both  previous and  following contexts for each input token.  Attention mechanisms\cite{bahdanau2014neural} use learned weights to help the network ``pay attention" to tokens that are more important than others in a context.

\citet{simple_ict}'s SIMPLE extended basic 1980s' style feedforward neural network with  state-of-the-art SE's optimizer DODGE~\cite{agrawal2019dodge} to automatically select the preprocessors (normalizer, binarizer, etc) and the neural network model's hyperparamters (num\_layers, num\_units\_in\_layer, batch\_size).  SIMPLE outperformed DeepTriage and other non-neural network methods from \citet{marks2011studying} and \citet{guo2010characterizing}.

SIMPLE is employed as the SOTA solution for predicting issue close time. \citet{simple_ict} only compared solutions by the accuracy metric. In order to ensure the generalizability of our proposed solution, we also compared different methods with metrics from static warning analysis in RQ2 (FAR, recall, and AUC). Hence,
Table \ref{tab:rq3} reports the comparison of our proposed method's FRUGAL, the SOTA's SIMPLE,  and the baseline unsupervised learners (CLA \& CLAFI+RF) across accuracy, FAR, recall, and AUC. We observe:

\bi
\item The unsupervised learners CLA/CLAFI+RF performed   worst as it's default behavior uses  clustering based on the median of the data (which may not apply for all   data). While CLAFI+RF performed better than CLA in static code warnings,  that effect was not seen here (i.e.   what works for one domain may not work for another). Additionally, on average, CLA  underperformed SIMPLE by approximately  1\%, 4\%, 4\%, 5\%, 5\%, and 11\% in FAR, precision, f1, recall, AUC,  and accuracy  without access to the train data. Altogether, both points indicate promising areas for tuning configurations of unsupervised learners.  

\item \revised{FRUGAL outperforms the SOTA's SIMPLE as FRUGAL wins in recall, precision, f1, AUC, and FAR while drawing in accuracy. FRUGAL, on average, improves relative SOTA's  precision,  AUC, f1, and recall by 9\% 14\%, 30\%, and 52\% respectively while reducing FAR by \revised{94\%} relatively.}   

\item In term of labelling efforts, CLA is label-free, FRUGAL costs \revised{2.5\%}, and the \citet{simple_ict}'s method costs 100\% because FRUGAL and the SOTA need \revised{2.5\%} and 100\% of the data labelled to execute.  

\ei 
\vspace{5pt}

\begin{RQ}{\normalsize{In summary, our answer to RQ3 is:}} 
FRUGAL exceeds both standard unsupervised learner CLA and the SOTA SIMPLE (EMSE'20 \cite{simple_ict} which outperformed a decade of research including  ICSE'10~\cite{giger2010predicting}, PROMISE'11~\cite{marks2011studying}, MSR'16~\cite{Kikas16},  COMAD'19~\cite{mani2019deeptriage}) in predicting issues close time. FRUGAL requires only \revised{2.5\%} of the train data to be labelled when being compared against unsupervised learning while using \revised{97.5\%} less information than the SOTA tuned deep learning method. Hence, FRUGAL is not only effective in static warning analysis, but also in issue close time prediction. The success in both areas let this study hypothesizes that other areas of SE may also benefit from FRUGAL.
\end{RQ}




\section{Threats of Validity}\label{tion:threats}

There are several validity threats~\cite{feldt2010validity} to the design of this study. Any conclusion made from this work must be considered with the following issues in mind:

\textbf{Conclusion validity} focuses on the significance of the treatment. To enhance   conclusion validity, we run experiments on 12 different target projects across stratified sampling (25 runs)  and find that our proposed method always performed better than the state-of-the-art approaches. More importantly, we apply a similar statistical testing of Cohen'd as the SOTA work~\cite{simple_ict, intrinsic_static} from the two domains to obtain fair comparison.  In addition, we have taken into generalization issues of single evaluation metrics (e.g., recall and precision) into consideration and instead evaluate our methods on metrics that aggregate multiple metrics like AUC while being effort-aware via cost.  As future work, we plan to test the proposed methods with additional analyses that are endorsed within SE literature (e.g., P-opt20~\cite{tu2020better}) or general ML literature (e.g., MCC~\cite{mcc_metrics}).

One of the possible explanations for the simple effectiveness of both binary split of the output space (CLA/CLAFI+ML/FRUGAL's centrality) and \revised{2.5\%} labelled train data requirement is highly due to the intrinsic dimensionality. \citet{levina2004maximum}
  argued that many datasets embedded in high-dimensional spaces can be compressed without significant information loss (similar to the PCA method~\cite{MACKIEWICZ1993303}).  To compute Levina's intrinsic dimensionality, a 2-d plot is created where the x-axis shows $r$; i.e. the radius of two configurations while the y-axis shows $C(r)$ as the number of configurations after spreading out some distance $r$ away from any of $n$ data instances:  
  
\begin{equation}
y = C(r) = {\frac{2}{n(n-1)}\sum\limits_{i=1}^n \sum\limits_{j=i+1}^n I\left[\lVert x_i, x_j \rVert < r \right]}
\end{equation}
The maximum slope of  $\ln C(r)$ vs. $\ln r$ is then reported as the intrinsic dimensionality, $D$.
Note that
$I[\cdot]$ is the indicator function (i.e., $I[x] = 1$ if $x$ is true, otherwise it is 0);
$x_i$ is the $i$th sample in the dataset. Applying this calculation to the 12 datasets of two domains (reports in Table \ref{tab:rq4}), we found the intrinsic or latent dimensionality ($D$) of our data is very low (median around one, no more than three).  Agrawal et al.'s DODGE~\cite{agrawal2019dodge} is the SOTA optimizer for SE, DODGE executes by binary splitting the tuning  space, each  chop moves in the bounds for numeric choices
by half the distance from most distant value to the value that produced the ``best'' performance. According to Agrawal et al., DODGE's effectiveness roots in how  the performance score generated from SE data can be divided into a few regions (low dimensional). FRUGAL's central function of binary splitting is similar to DODGE as FRUGAL compresses the data dimensions (features) via aggregated percentile $C$ and survey the whole space by varying $C$ ($\{5\%$ to $95\%$ increments by $5\%\}$). Menzies et al.~\cite{menzies07} and Hindle et al.~\cite{hindle2012naturalness} also reported on how several SE data are low dimensional and the benefits from building effective tools from such data. This work extends those findings: the labelling efforts to commission to tools building  can be reduced greatly because of the low dimensionality of SE data. 

\textbf{Internal validity }focuses on how sure we can be that the treatment caused the outcome. To enhance   internal validity, we heavily constrained our experiments to the same dataset, with the same settings, except for the treatments being compared.

\newcolumntype{Y}{>{\centering\arraybackslash}X}

\begin{table}[!t]
\small
\caption{Summary of intrinsic dimensions ($D$) of this study's 12 datasets from \citet{levina2004maximum}.}
\vspace{-5pt}
\centering
\setlength\tabcolsep{2pt}
\renewcommand{\arraystretch}{1.05}

\begin{tabularx}{\linewidth}{P{.03\linewidth}P{.055\linewidth}|P{.055\linewidth}|P{.055\linewidth}|P{.055\linewidth}|P{.055\linewidth}|P{.055\linewidth}|P{.055\linewidth}|P{.055\linewidth}|P{.055\linewidth}|P{.08\linewidth}|P{.08\linewidth}|P{.07\linewidth}}
\midrule
 & \multicolumn{9}{c}{\textbf{Static Code Warnings}} & \multicolumn{3}{c}{\textbf{Issue Close Time}} \\ 
 \cmidrule(lr){2-10} \cmidrule(l){11-13}

   \textbf{} &  \textbf{\rotatebox[origin=c]{90}{Derby}} & \textbf{\rotatebox[origin=c]{90}{Mvn}} & \textbf{\rotatebox[origin=c]{90}{Lucene}} & \textbf{\rotatebox[origin=c]{90}{Phoenix}} & \textbf{\rotatebox[origin=c]{90}{Cass}} & \textbf{\rotatebox[origin=c]{90}{Jmeter}} & \textbf{\rotatebox[origin=c]{90}{Tomcat}} 
   & \textbf{\rotatebox[origin=c]{90}{Ant}}
   & \textbf{\rotatebox[origin=c]{90}{Commons}} &  \textbf{\rotatebox[origin=c]{90}{Chromium}} &  \textbf{\rotatebox[origin=c]{90}{Firefox}}  &  \textbf{\rotatebox[origin=c]{90}{Eclipse}} \vspace{3pt}\\

\midrule

\multicolumn{1}{c|}{\textbf{$D$}} & 0.78 & 1.10 & 0.15  & 0.62  &  1.94 & 1.54 & 0.73  & 0.82 &   1.04       &  1.95 \hspace{2pt}  &   \hspace{1pt} 2.10 &    \hspace{1pt} 1.9 \\

\bottomrule
\end{tabularx}
\label{tab:rq4}
\vspace{-20pt}
\end{table}

\textbf{Construct validity }focuses on the relation between the theory behind the experiment and the observation. To enhance construct validity, we compared solutions with and without our strategies in Table~\ref{tab:rq2} and \ref{tab:rq3} while showing that both components (unsupervised learning with CLA/CLAFI+ML~\cite{nam2015clami} and tuned semi-supervised method of FRUGAL) and in various amounts of labelled data required for the proposed mehtod  to improve the overall performance. \revised{Moreover, we also benchmarked our solution with the SOTA's solution that is trained on the same $L=2.5\%$ to ensure that our proposed solution's effectiveness is not due to random sampling of the data.} However, we only show that with our default parameters settings of random forest learner. The performance can get even better by tuning the parameters, employing different learners (e.g., deep learners), and introducing a variety of data preprocessors (e.g., synthetic minority over-sampling or SMOTE that is known to help with imbalanced datasets \cite{chawla2002smote,agrawal2018better} like our static code  warnings domains). We aim to explore these in our future work.

\textbf{External validity} concerns how widely our conclusions can be applied. In order to test the generalizability of our approach, we always kept a project as the holdout test set and never used any information from it in training. Moreover, we have validated our proposed method on two important software analytics \revised{domains}: adoptable static code warnings identification and issues close time prediction. Our experiments with default CLA/CLAFI+ML~\cite{nam2015clami} demonstrates the danger of treating
all data with the state-of-the-art method, especially when switching domain (from defect prediction to issue close time prediction and adoptable static code warning identification).

\section{Conclusion and Future Work} \label{sec:conclusion}

There is much recent advance for software analytics research with automated and semi-automated methods. However, these  methods are built on a sufficiently large amount of data labelled. Generating such labels can be labor-intensive and expensive (as discussed in \S\ref{sec:motivations}). Such requirement can introduce barrier for entering new research domains (e.g., the success of open-source projects).  In order to reduce the label famine and human effort, FRUGAL is recommended. FRUGAL tunes the state-of-the-art unsupervised learner from defect prediction (CLA/CLA+ML/CLAFI+ML) and it's corresponding percentile parameter $C$ in the grid search manner while validating on only \revised{\revised{2.5\%}} of the labelled data. Our findings include:


\be
\item Unsupervised Learners without access to the train data's labels performed approximately 10\% less than the SOTA methods on average. The results are promising but still not effective enough. 
\item FRUGAL performed similarly to the SOTA adoptable static code warning identifier while surpassing the SOTA issue close time predictor with \revised{97.5\%} less information.  
\item FRUGAL reduced the labelling efforts needed for the software analytics tools by \revised{97.5\%}. Simply, FRUGAL is \revised{40(100\% / 2.5\%)} times cheaper than the SOTA methods in issue close time and static code warnings analysis areas. 
\item The success of FRUGAL for the two \revised{domains} here suggests that many more domains in software analytics could benefit from unsupervised learning. As mentioned above, those
benefits include the ability to commission new models with less human efforts and costs. By restricting human involvement in the process, we also reduced erroneous labels that can cascade to the whole research community since human are still error-prone (Yu et al.~\cite{jitterbug} found 98\% of the false-positive labels within \citet{maldonado2015detecting} were actually true-positive labels).
\item Overall, our proposed method restated the benefit in exploring low dimensional SE data~\cite{hindle2012naturalness, agrawal2019dodge, menzies07, intrinsic_static, simple_ict} and extended their findings that the labelling efforts can be reduced greatly because of the low dimensionality of SE data.
\ee

That said, FRUGAL still suffers from the validity threats discussed in \S\ref{tion:threats}. To further reduce those threats and to move forward with this research, we propose the following future work:
\bi
\item \revised{Comparing FRUGAL to other semi-supervised learning methods within software engineering, e.g., NSGLP~\cite{nsglp}.}
\item Test whether replacing the Random Forest model in FRUGAL  with a deep learning model will further improve its performance.
\item Explore non-SE or high-dimensional SE data with FRUGAL to see if our current conclusions still hold. 
\item Apply non-trivial hyper-parameter tuning (e.g., DODGE~\cite{agrawal2019dodge} or FLASH~\cite{nair2017flash}) on various data preprocessors and machine learners with FRUGAL to test whether tuning can further improve the performance.
\item
Extend the work to other software engineering domains (e.g., security~\cite{5463340}, technical debts~\cite{maldonado2015detecting}, software configurations~\cite{Estublier}, etc) and compare it with other state-of-the-art methods which continue to appear.
\ei

\section*{Acknowledgements}
This work was partially funded  
 by an NSF CISE Grant \#1931425.
\balance
\bibliographystyle{plainnat}

\clearpage
\appendix

\end{document}